\documentclass[prd,tightenlines,nofootinbib,amsfonts,amssymb,amsmath,color,11pt]{revtex4}
\pdfoutput=1
\usepackage{graphicx}
\usepackage{epsfig}
\usepackage{rotate}
\usepackage{amsmath}
\usepackage{amssymb}
\usepackage{amsfonts}
\usepackage{enumerate}
\usepackage{afterpage}
\usepackage{color}
\usepackage{amscd}
\usepackage{hyperref}

\def\be{\begin{equation}}
\def\ee{\end{equation}}
\def\bea{\begin{eqnarray}}
\def\eea{\end{eqnarray}}

\newcommand{\no}{\noindent}
\newcommand{\nb}{\nonumber}

\newcommand{\de}{\partial}

\renewcommand\a{\alpha}

\renewcommand\b{\beta}

\newcommand\m{\mu}
\newcommand\n{\nu}

\newcommand\Ga{\Gamma}
\newcommand\Om{\Omega}
\newcommand\Ost{Ostrogradsky }

\begin{document}

\title{Extended Scalar-Tensor Theories of Gravity}

\author{Marco Crisostomi$^a$, Kazuya Koyama$^a$ and Gianmassimo Tasinato$^{b}$}
\affiliation{
	$^a$Institute of Cosmology and Gravitation,~University of Portsmouth, Portsmouth, PO1 3FX, UK\\
	$^b$Department of Physics, Swansea University, Swansea, SA2 8PP, UK
}

\hskip1.5cm
\begin{abstract}
\no  We study new consistent scalar-tensor theories of gravity recently introduced by Langlois and Noui with potentially interesting cosmological applications. We derive the conditions for the existence of a primary constraint that prevents the propagation of an additional dangerous mode associated with higher order equations of motion. We then classify the most general, consistent scalar-tensor theories that are at most quadratic in the second derivatives of the scalar field. In addition, we investigate the possible connection between these theories and (beyond) Horndeski
through conformal and disformal transformations. Finally, we point out that these theories can be associated with new operators in the effective field theory of dark energy, which might open up new possibilities to test dark energy models in future surveys.
\end{abstract}

\maketitle

\section{Introduction}
\no Scalar tensor theories of gravity play an essential role for building models of inflation and dark energy. The most general scalar-tensor theory leading to second order equations of motion, the theory of Horndeski \cite{Horndeski:1974wa, Deffayet:2011gz}, propagates three degrees of freedom (dof) in the vacuum. When coupled with matter, this set-up provides a convenient framework for parametrising linear cosmological perturbations and for developing tests of General Relativity using cosmological observations (see e.g. \cite{Koyama:2015vza} for a review). On the other hand, it has been recently pointed out \cite{Gleyzes:2014dya, Gleyzes:2014qga, Gao:2014soa} that there are scalar-tensor theories more general than Horndeski, which are characterised by higher order equations of motion, but which nevertheless propagate three degrees of freedom in the  vacuum. This property is ensured by the existence of  constraints that forbid the propagation of additional, dangerous modes. Such theories are  dubbed `beyond Horndeski'   \cite{Gleyzes:2014dya, Gleyzes:2014qga}.
 
The existence of these theories raises various broad questions that we consider in this paper:
\begin{itemize}
\item  What is the {\it most general}, consistent theory of gravity coupled with a scalar field? Beyond Horndeski is an explicit example of a consistent generalisation of Horndenski, but it is not necessarily the most general one. In fact, \cite{Langlois:2015cwa,Langlois:2015skt} provided explicit examples of theories more general than beyond Horndeski. They study degeneracy conditions for the kinetic matrix associated with Lagrangians that are at most quadratic in the scalar second derivatives. In the present work, we re-derive the results of \cite{Langlois:2015cwa} using a method that 
provides the conditions for the existence of a primary constraint starting from the conjugate momenta of the scalar and tensor fields.  Our method is equivalent to the one developed in \cite{Langlois:2015cwa} to impose the degeneracy conditions for the kinetic matrix.
In the present work we limit our attention to theories quadratic in second derivatives of the scalar and
we dub them {\it extended scalar-tensor} theories of gravity, in brief EST theories, and classify them carefully.  
  
\item Is there any connection between EST theories and the original Horndeski, or beyond Horndeski actions? It is known that generalised disformal transformations allow to generate beyond Horndeski Lagrangians, starting from Horndeski's \cite{Gleyzes:2014qga,Crisostomi:2016tcp}.    In this work we study which EST set-ups can be obtained by generalised conformal and/or disformal  transformations of Horndeski and beyond Horndeski actions. See  \cite{Domenech:2015hka,Motohashi:2015pra,Fujita:2015ymn,Domenech:2015tca,Bettoni:2013diz,Zumalacarregui:2013pma,Sakstein:2015jca} for works studying related aspects of disformal transformations in scalar-tensor theories and applications to cosmology.

\item Are there distinctive  phenomenological consequences associated with  EST theories? We make progresses in answering this question by considering two phenomenological aspects of EST theories. First, we  study which classes among the EST theories admit a healthy Minkowski limit: a necessary condition when considering applications to weakly gravitating relativistic systems. Then, we turn to cosmology and examine the possible relevance of EST theories for dark energy, using the language of the Effective Field Theory (EFT) of dark energy \cite{Gubitosi:2012hu,Bloomfield:2012ff}. Such approach provides a powerful tool for connecting general features of scalar-tensor theories to EFT operators that control the evolution of cosmological perturbations. We  show that EST theories can be associated with novel EFT operators for dark energy, which are absent in Horndeski and beyond Horndeski theories. Hence, these theories can have potentially distinctive new observational consequences for dark energy model building.
\end{itemize}
 
\no The building blocks for our analysis are general scalar-tensor Lagrangian densities that are at most quadratic on second derivatives of the scalar field. Without loss of generality, we can express such Lagrangian densities as linear combinations of the following pieces \cite{Langlois:2015cwa}
\be
{\cal L}_{\text{tot}}\,=\,\sum_{i=1}^5{\cal L}_i+{\cal L}_R \,,
\ee
where
\begin{align}
{\cal L}_1 [A_1] &= A_1(\phi,\,X) \phi_{\mu \nu} \phi^{\mu \nu} \,,
\label{A1}
 \\
{\cal L}_2 [A_2] &= A_2(\phi,\,X) (\Box \phi)^2 \,, 
\label{A2}
 \\
{\cal L}_3 [A_3] &= A_3(\phi,\,X) (\Box \phi) \phi^{\mu} \phi_{\mu \nu} \phi^{\nu} \,, 
\label{A3}
\\
{\cal L}_4 [A_4] &= A_4(\phi,\,X)  \phi^{\mu} \phi_{\mu \rho} \phi^{\rho \nu} \phi_{\nu} \,, 
\\
{\cal L}_5 [A_5] &= A_5(\phi,\,X)  (\phi^{\mu} \phi_{\mu \nu} \phi^{\nu})^2\,, 
\end{align}
while
\begin{equation}
{\cal L}_R[G] = G(\phi,\,X) R \,,
\label{R}
\end{equation}
is a non-minimal coupling with gravity\footnote{This is indeed the most general non-minimal coupling with gravity involving quadratic powers of the velocities. Couplings involving the Riemann and Ricci tensor can be reduced to this one through symmetric reasons and integrations by parts.}. 
We defined $\phi_{\mu} = \de_{\mu} \phi, \phi_{\mu \nu} = \nabla_{\mu} \nabla_{\nu} \phi$ and $X = \phi^{\mu} \phi_{\mu}$.  The functions $G$, $A_i$ are arbitrary functions of $\phi$ and  $X$: for simplicity, we will only consider them to be functions of $X$ in our
analysis.

This general way of combining Lagrangians ${\cal L}_i$ includes the theory of quartic Horndeski, described by  
\begin{equation}
{\cal L }_H  = G\, R - 2 G_X \left[ ( \Box \phi)^2 - \phi_{\mu \nu} \phi^{\mu \nu} \right] \,, 
\label{H}
\end{equation}
(where  $G_X = \partial G/\partial X$),
as well as quartic beyond Horndeski, given by 
\begin{equation}
{\cal L}_{BH} =  F(\phi,\, X) \left[
X \left( (\Box\phi)^2-\phi_{\mu \nu} \phi^{\mu \nu} \right)
-2 \left( \Box \phi\, \phi^{\mu}
\phi_{\mu\nu}\phi^{\nu} 
- \phi^{\mu} \phi_{\mu \rho} \phi^{\rho \nu} \phi_{\nu} \right) \right].
\label{BH}
\end{equation} 

This paper is organised as follows. In section II we present our formalism to identify the conditions to ensure the existence of a primary constraint in the system described by the Lagrangians Eqs.~(\ref{A1})$-$(\ref{R}). In section III we classify the consistent EST theories with minimal and non-minimal couplings to gravity.
In section IV we identify the class of models that can be obtained from Horndeski and beyond Horndeski theories  by generalised conformal and disformal transformations. In section V we discuss phenomenological applications of the new theories. We first identify theories that admit a healthy Minkowski limit, then we identify new operators for the EFT of dark energy. Section VI is devoted to our conclusions.

\section{Primary constraints in EST theories}
\no The set of Lagrangians ${\cal L}_i$ presented in the Introduction leads to higher order equations of motion. Hence they are generally associated with the \Ost instabilities, unless there are constraints that forbid the propagation of additional dangerous modes.
In this section we provide a general tool to identify the combinations of Lagrangians characterised by the presence of an extra primary constraint.
With the word ``extra'' we mean a constraint necessary to eliminate the \Ost mode, meanwhile we assume that the first class constraints associated with the diffeomorphism (diff.) invariance still persist.
We are aware that a primary constraint is necessary but not sufficient to remove a propagating dof, however, due to the Lorentz invariance of the theories, we expect that there is always an associated secondary constraint generated by a second class primary constraint.  
Indeed \cite{Langlois:2015skt} showed that this was the case.
Primary constraints exist when, passing to the Hamiltonian formalism,
all the velocities {\it cannot} be expressed in terms of the fields and their conjugate momenta.
This translates to relations (constraints) between the fields and momenta that need to be added to the canonical Hamiltonian through Lagrangian multipliers.  See also e.g. \cite{Deffayet:2015qwa,Lin:2014jga,Gao:2014fra} 
for works studying the Hamiltonian structure of scalar tensor theories.

\subsection{Kinetic terms and conjugate momenta}
 
\no In order to identify the kinetic terms and carry out the analysis of constraints we need to separate space and time, performing a 3+1 decomposition as in \cite{Langlois:2015cwa,Crisostomi:2016tcp}. We introduce the time vector flow $t^\mu\,=\,\partial/\partial t$ decomposed as $t^\mu\,=\,N\,n^\mu+N^\mu$, where $n^\mu$ is the time-like unit normal vector to the $t=$ constant  hypersurface, $N$ the lapse function and $N^\mu$ the shift vector orthogonal to the normal vector. The constant time hypersurface is then characterised by the normal vector $n^{\mu}$, the 3D metric $h^{\mu}_{\,\,\nu} =\delta^{\mu}_{\nu} + n^{\mu} n_{\nu}$ and the extrinsic curvature
\be\label{ext-def}
K_{\mu \nu}\,=\,\frac{1}{2 N}\left( \dot{h}_{\mu\nu}-D_{(\mu} N_{\nu)} \right) \,,
\ee 
where `dot' is the Lie derivative respect to $t^\mu$\,, $D_{\mu}$ is the 3D covariant derivative and the parenthesis $(\dots)$ on the indices denote symmetrisation with no $1/2$ factor.  

To study the dynamics of this kind of Lagrangians (i.e. with second derivatives in the action), it is useful to identify, with the help of a Lagrangian multiplier,
\be
\partial_\mu \phi 
 \equiv A_{\mu}\,,
\ee
with $A_\mu$ an auxiliary  vector field. We   decompose $A_{\mu}$ into the normal and transverse components with respect to the aforementioned hypersurface:
\begin{equation}
	A_{\mu} = -A_* n_{\mu} + \hat{A}_{\nu} h^{\nu}_{\,\mu}\,.      
	\label{hatA}
\end{equation}
The covariant derivative of $A_{\mu}$ can be decomposed into various pieces depending on the derivatives of its components and of the metric:
\bea \label{dmuanuc}
\nabla_{\mu} A_{ \nu} = 
D_{\mu} \hat A_{\nu}-A_*\, K_{\mu\nu}+n_{(\mu} \left( K_{\nu) \rho } \hat{A}^\rho-D_{\nu)}A_*\right)
+\,n_\mu n_\nu \left(V_*- \hat{A}_\rho \,a^\rho \right) \,,
\eea
where $a^\mu=n^\nu \,\nabla_{\nu} \,n^\mu$ is the acceleration vector and 
\be
V_* \equiv n^{\mu} \nabla_{\mu} A_* = \frac{1}{N} \left( \dot{A}_* -N^\mu D_\mu A_* \right)\,. \label{Vstar}
\ee
In Eq.~(\ref{dmuanuc}) time derivatives appear only for the three dimensional metric $h_{\m\n}$ (inside the extrinsic curvature) and for the component $A_*$ (inside $V_*$). $V_*$ plays for $A_*$ the same role that $K_{\m\n}$ plays for~$h_{\m\n}$.
 
In \cite{Crisostomi:2016tcp} we introduced a novel way to perform the Hamiltonian analysis of the system, 
allowing us to keep a covariant structure even when the space-time is characterised by a special foliation.
%
 It consists on working directly with the extrinsic curvature and $V_*$, instead of using the true velocities~$\dot{h}_{\m\n}$ and $\dot{A}_*$, identifying the terms in the action containing $V_*$ and $K_{\mu\nu}$ as the kinetic contributions to it. 
An advantage of this procedure is that the Lagrangian densities do not depend explicitly on the lapse and shift functions: such quantities are indeed implicitly included in $K_{\,\,\nu}^\mu$ and $V_* $. 
We define therefore the conjugate momenta for $A_*$ and $h_{\mu \nu}$ accordingly:
\be
 \pi_* \equiv \frac{1}{\sqrt{-g}} \frac{\delta {L}}{\delta V_*} \,, \qquad
 \pi^{\a}_{\m} \equiv \frac{1}{\sqrt{-g}}
 \frac{\delta {L}}{\delta K^{\mu}_{\alpha}} \,.
 \ee
Notice that this definition is slightly different from the usual one due to the presence of the factor $1/\sqrt{-g}$, which completely removes the lapse from the expressions. For the purpose of proving the existence of primary constraints, this difference is not important.

The conjugate momenta associated
with each Lagrangian ${\cal L}_i$ are given by \\
\begin{itemize}
\item ${\cal L}_1$:
\bea
\pi_* &= & 2 V_* \,, \\
\pi^{\alpha}_{\mu} &= & 
2 \left(A_*^2 K^{\alpha}_{\mu} - \hat{A}_{\mu} \hat{A}^{\nu} K^{\alpha}_{\nu} - \hat{A}_{\nu} \hat{A}^{\a} K^{\nu}_{\mu} \right) \,, 
\eea
\item ${\cal L}_2$:
\bea
\pi_* &= & 2 \left(A_* K +  V_*\right) \,, \\
\pi^{\alpha}_{\mu} &= & 2 A_* h^{\alpha}_{\mu} \left(A_* K  + V_*\right) \,, 
\eea
\item ${\cal L}_3$:
\bea
\pi_* &= & - A_* \left(\hat{A}_{\mu} \hat{A}^{\a} K^\mu_\a + A_*^2 K + 2 A_* V_*\right) \,, \\ 
\pi^{\alpha}_{\mu} &= & 
- A_* \hat{A}_{\mu} \hat{A}^{\alpha} \left(A_* K + V_*\right) - A_*^2 h^{\alpha}_{\mu} \left(\hat{A}_{\nu} \hat{A}^{\b} K^\nu_\b +  A_* V_* \right) \,, 
\eea
\item ${\cal L}_4$:
\bea
\pi_* &= & -2 A_* \left(\hat{A}_{\mu} \hat{A}^{\a} K^\mu_\a +  A_* V_*\right) \,, \\
\pi^{\alpha}_{\mu} &= & 
- 2 \hat{A}_{\mu} \hat{A}^{\alpha} \left( \hat{A}_{\nu} \hat{A}^{\b} K^\nu_\b + A_* V_*\right) \,,
\eea
\item ${\cal L}_5$:
\bea
\pi_* &= & 2 A_*^3 \left(\hat{A}_{\mu} \hat{A}^{\a} K^\mu_\a +  A_* V_*\right) \,, \\
\pi^{\alpha}_{\mu} &= & 
2 A_*^2 \hat{A}_{\mu} \hat{A}^{\alpha} \left( \hat{A}_{\nu} \hat{A}^{\b} K^\nu_\b + A_* V_* \right) \,,
\eea
\item ${\cal L}_R [G]$:
\bea
\pi_* &= & 4 G_X A_* K \,, \\
\pi^{\alpha}_{\mu} &= & 2 G \left(  K_\mu^\a - K h_\mu^\a\right) + 4 G_X \left[  K \hat{A}_{\mu} \hat{A}^{\alpha} + h_\mu^\a \left( \hat{A}_{\nu} \hat{A}^{\b} K^\nu_\b + A_* V_* \right) \right] \,.
\eea
\end{itemize}
Notice that we only keep terms that involve the extrinsic curvature $K_{\mu \nu}$ and $V_*$, as these are  the only relevant ones for the construction of primary constraints.
Indeed in this paper we are interested in showing the existence of primary constraints and not in their exact form (that is instead needed to study their evolution); hence the relations that we will give are exact only in the momenta but not in the fields.

\subsection{Constructing the theories: a general tool}

\no The most general scalar primary constraint involving the momenta $\pi_*$ and $\pi^{\alpha}_{\mu}$ takes the following form\footnote{Indeed there are only two scalar combinations that can be constructed out of $\pi_{\mu}^{\alpha}$, i.e. its trace and the projection along $\hat{A}$.}
\be
\left(a\, h_{\alpha}^{\mu} + b\,  \hat{A}_{\alpha} \hat{A}^{\mu}\right) \pi_{\mu}^{\alpha} + c\,  \pi_* + d \approx 0 \,, \label{genprimary}
\ee
where  $a,\, b,\, c$ and $d$ are functions of the fields and the symbol  
``$\approx$'' indicates weak equality, i.e. equality on the phase space determined by constraints.
For simplicity, we do not provide the expression for $d$ in (\ref{genprimary}), as it is not relevant  for our purposes.

For instance, the Lagrangians  ${\cal L}_2, \,{\cal L}_3,\, {\cal L}_4$ and ${\cal L}_5$ enjoy the following primary constraint
\begin{align}
{\cal L}_2:\qquad &  \left(a\, h_{\alpha}^{\mu} + b\,  \hat{A}_{\alpha} \hat{A}^{\mu} \right) \pi^{\alpha}_{\mu} - A_* \left(3\, a + b\, \hat{A}^2 \right) \pi_* \approx 0 \,, \label{A2cost} \\
{\cal L}_3:\qquad &  \left( h_{\alpha}^{\mu} - \frac{\hat{A}^2 - 3 A_*^2}{A^2 \hat{A}^2}\, \hat{A}_{\alpha} \hat{A}^{\mu} \right) \pi^{\alpha}_{\mu} - \frac{2 A_* \hat{A}^2}{A^2}\, \pi_* \approx 0 \,,\label{const3} \\
{\cal L}_{4,5}:\qquad &  \left(a\, h_{\alpha}^{\mu} + b\,  \hat{A}_{\alpha} \hat{A}^{\mu} \right) \pi^{\alpha}_{\mu} - \frac{\hat{A}^2}{A_*} \left( a + b\,\hat{A}^2 \right) \pi_* \approx 0 \,, \label{A45cost}
\end{align}  
so, when isolated, these Lagrangians propagate less than four degrees of freedom\footnote{Notice that for ${\cal L}_{2,4,5}$ a primary constraint exists without the necessity of a tuning between $a$ and $b$, instead in ${\cal L}_{3}$ this is needed. This implies that, when these Lagrangians are joined together, in order to form a unique primary constraint, the values of $a$ and $b$ have to tune to the ones in (\ref{const3}).}.

\smallskip

More in general, in order to find the theories possessing the kind of constraint in (\ref{genprimary}), there is a  simple requirement to impose:
the ratios between the coefficients of the velocities in $\pi_*$ and $(a\, h_{\alpha}^{\mu} + b\,  \hat{A}_{\alpha} \hat{A}^{\mu}) \pi_{\mu}^{\alpha}$ should be the same.
For the Lagrangians (\ref{A1})$-$(\ref{R}) the velocities appear only in three forms:
\begin{equation}
V_*\,, \qquad K \,, \qquad \hat{A}_{\mu} \hat{A}^{\a} K^\mu_\a \,,
\end{equation}
therefore there are two conditions to impose:
\bea
\left. \frac{\text{Coeff.}(V_*)}{\text{Coeff.}(K)} \right|_{\pi_*} &=& \left. \frac{\text{Coeff.}(V_*)}{\text{Coeff.}(K)} \right|_{(a\, h_{\alpha}^{\mu} + b\,  \hat{A}_{\alpha} \hat{A}^{\mu}) \pi_{\mu}^{\alpha}} \,, \\[2ex]
\left. \frac{\text{Coeff.}(\hat{A}_{\mu} \hat{A}^{\a} K^\mu_\a)}{\text{Coeff.}(K)} \right|_{\pi_*} &=& \left. \frac{\text{Coeff.}(\hat{A}_{\mu} \hat{A}^{\a} K^\mu_\a)}{\text{Coeff.}(K)} \right|_{(a\, h_{\alpha}^{\mu} + b\,  \hat{A}_{\alpha} \hat{A}^{\mu}) \pi_{\mu}^{\alpha}} \,.
\eea
For a generic combination of ${\cal L}_i$ ($i=1, ... ,5)$ and ${\cal L}_R$,
these two requirements lead to very specific conditions on the free functions involved, $A_i(X)$ and $G(X)$. 
In practice we can solve one of the conditions for $a$ (or $b$) and putting the result in the other one we get an equation polynomial in~$A_*$. 
In order to obtain Lorentz invariant solutions for $A_i(X)$ and $G(X)$, this equation needs to be satisfied at every order in $A_*$.
 This gives the following three conditions
\bea
&& (A_1 + A_2) \left[G\, X (2 A_1 + A_4\, X + 4 G_X) - 2 G^2 - 8 G_X^2 X^2\right] = 0
\label{first}
\,, \\[2ex]
&& 4 X \left[G \left(A_3^2 - 4 A_2 A_5\right) - 2 A_1 A_3 G_X + A_1^2 A_3 \right] - 4 \left(2 A_1^2 A_2 + 2 A_1 A_3 G + A_1^3 + 2 A_5 G^2\right) \nb \\
&& + A_1 X^2 \left[4 A_5 (A_1 + 3 A_2) - 3 A_3^2\right] +16 G_X \left(2 A_1 A_2 + A_1^2 + A_3 G\right) -16 G_X^2 (A_1 + 2 A_2) = 0 \,, 
\label{second}
\\[2ex]
&& X^2 \left[A_3^2 G - 4 \left(A_1 (3 A_2 A_4 - 4 A_3 G_X + A_5 G) + A_1^2 A_4 + A_2 A_5 G\right)\right] \nb \\
&& - 4 X \left[A_1 (3 A_3 G - 20 A_2 G_X) + A_1^2 (6 A_2 - 4 G_X) + 2 A_1^3 - 4 A_2 A_4 G + 16 A_2 G_X^2 + 2 A_3 G\, G_X \right] \nb \\
&& + 4 G \left[(A_1 - 2 G_X) (A_1 + 4 A_2 + 6 G_X) + 2 G (A_3 + A_4)\right] = 0 \,.
\label{third}
\eea 
These conditions agree with those obtained in
Ref.~\cite{Langlois:2015cwa} demanding that the kinetic matrix is degenerate.

  
\section{EST Theories: the classification} 

\subsection{Minimally coupled theories}\label{sub-mct} 

\no In this subsection we classify solutions for the 
constraint equations \eqref{first}-\eqref{third} in the absence of non-minimal coupling to gravity. By assuming $G(X)=0$, the conditions to have a primary constraint become simpler
\begin{align}
&A_1 (A_1 + 3 A_2)(2 A_1 + A_4 X) =0 \,, \\
&A_1 \left[X^2 \left(3 A_3^2 - 4 A_5 (A_1 + 3 A_2) \right)+ 4 A_1 (A_1 + 2 A_2) - 4 A_1 A_3 X \right] =0 \,.
\end{align}
From the first condition, we find that there are three branches of the solutions. We study these three branches in turn. 

\hspace{1cm}\\
$\blacktriangleright$ {\bf (M-I):  $A_4 = - 2\,A_1/X$} \\

In this case, the second condition can be used to express one of the functions $A_2, A_3$ or $A_5$ in terms of $A_1$ and the others.
If $A_2 \neq - A_1/3$ then $A_5$ can be determined as 
\begin{equation}
A_5 = \frac{4 A_1 \left( A_1 + 2 A_2 \right) - 4 A_1 A_3 X + 3 A_3^2 X^2}{4 \left(A_1 + 3 A_2\right) X^2}.
\label{MIa5}
\end{equation}
Thus there are three free functions in this case. The relation between  $b$ and $a$ is given by 
\be
b = \frac{2 A_1 + 4 A_2 - X A_3}{X \left(  2 A_2 + X A_3 \right)} \, a.
\ee
Note that for the following choice of parameters
\begin{equation}
A_3=  - \frac{2 A_2}{X}\,, 
\label{bhe}
\end{equation}
we have $a=0$ and $ A_5= (A_1+A_2)/X^2$. This case includes (but is not limited to)  beyond Horndeski theory: in this theory we have
\begin{equation}
A_2 = - A_1\,, \qquad A_3= - A_4 = \frac{2 A_1}{X}\,, \qquad A_5=0 \,, 
\end{equation}
where $A_1= - X F$.
This combination is special as it eliminates $V_*$ from $\pi_*$, thus there is no $V_*^2$ in the Lagrangian.

Let us make an aside, and point out that within this subclass of EST theories satisfying \eqref{bhe} there are Lagrangians admitting a particularly simple and elegant formulation. In  \cite{Crisostomi:2016tcp} we have shown that the theory of beyond Horndeski can be formulated using the projection tensor $P^{\alpha}_{\,\mu}$ on the constant scalar field hypersurface:
\begin{equation}
{\cal L}_{BH} = X F {\cal M}^{\alpha \beta}_{\mu \nu} \, \nabla_\alpha A^{\mu} \,\nabla_\b A^{\nu} \,, \qquad 
{\cal M}^{\alpha \beta}_{\mu \nu} = P^{\alpha}_{[\mu} P^{\beta}_{\nu]} \,, \qquad  
P^{\alpha}_{\,\mu} = \delta^{\alpha}_{\,\mu} - \frac{1}{X} A_{\mu} A^{\alpha} \,.
\end{equation} 
We can easily extend this theory to a more general one given by 
\begin{equation}
{\cal L} =  \, {\cal N}^{\alpha \beta}_{\mu \nu}\, \nabla_\alpha A^{\mu} \,\nabla_\b A^{\nu} , \qquad 
{\cal N}^{\alpha \beta}_{\mu \nu}\, =\, Q_1\, P^{\alpha}_{\mu} P^{\beta}_{\nu} + Q_2\, P^{\alpha}_{\nu} P^{\beta}_{\mu}.  \label{bbh-1}
\end{equation}
for arbitrary $Q_{1,2}(\phi,\,X)$. The existence of the primary constraint is guaranteed by the following property, which is due to the fact that ${\cal N}$ is built with projectors, and it is the same as the one we used in \cite{Crisostomi:2016tcp} for analysing beyond Horndeski: 
\begin{equation}
{\cal N}_{\mu\nu}^{\alpha \beta} \,A^\mu = {\cal N}_{\mu\nu}^{\alpha \beta} \,A_\alpha =0 \,. \label{propea}
\end{equation}
Notice that the previous property \eqref{propea} implies
\bea
A_*\, {\cal N}_{\mu\nu}^{\alpha \beta}\,n_\alpha n^\mu
&=& {\cal N}_{\mu\nu}^{\alpha \beta} \,\hat A_\alpha \,\hat A^\mu\,,
\\
A^2_*\, {\cal N}_{\mu\nu}^{\alpha \beta}\,n_\alpha n^\mu
&=& {\cal N}_{\mu\nu}^{\alpha \beta} \,\hat A_\alpha n^\mu\,.
\eea
By using these relations, it is easy to see that the conjugate momenta 
\bea
\label{cps}
\pi_*&=&2\,{\cal N}_{\mu\nu}^{\alpha \beta}\,n_\alpha n^\mu\,\nabla_\beta A^\nu,
\\
\hat A_\a \,\hat A^\mu\,\pi^\a_{\mu}
&=&2 \,{\cal N}_{\mu\nu}^{\alpha \beta}\,\left( -A_* \,\hat A_\alpha \,\hat A^\mu
+\hat A^2 n_\alpha \hat A^\mu+\hat A^2 \hat A_\alpha n^\mu
 \right)\,\nabla_\beta A^\nu,
 \label{cpk}
\eea
are proportional to each other, and there exists a constraint equation. Hence this theory satisfies the condition (\ref{bhe}), thus  providing a natural extension of beyond Horndeski theory. 

\hspace{1cm}\\
$\blacktriangleright$ {\bf (M-II) $A_2 = - A_1/3 $ }  \\

The second condition is satisfied if
\be
A_3 = \frac{2\,A_1}{3\,X} \,.
\ee
In this case $A_1$, $A_4$ and $A_5$ are free. The solution for $b$ and $a$ is given by 
\be
b = - \frac{a}{X}. 
\ee

\hspace{1cm}\\
$\blacktriangleright$ {\bf (M-III) $A_1=0$} \\

In this case, the second condition is automatically satisfied. Thus $A_2,A_3,A_4$ and $A_5$ are all free. This also means that ${\cal L}_2,{\cal L}_3,{\cal L}_4$ and ${\cal L}_5$ have a primary constraint individually as shown in Eqs. (\ref{A2cost}-\ref{A45cost}).
We can then easily show that any linear combination of $A_2,A_3,A_4$ and $A_5$ has the primary constraint given in equation (\ref{const3}) and 
\begin{equation}
b =  - \frac{\hat{A}^2 - 3 A_*^2}{A^2 \hat{A}^2} \, a.
\end{equation}

\subsection{Non-minimally coupled theories}

\no In this section we include a non-minimal coupling term \cite{Langlois:2015cwa}
\begin{equation}
{\cal L}_R[G] = G(\phi,\,X) R \,. 
\end{equation}
From the first condition (\ref{first}), there are two branches of solutions depending on $A_1 + A_2 =0$ or not. In the following discussions, we do not consider special cases where $G$ needs to be some specific function of $X$ for solving the equations. 

\hspace{1cm}\\
$\blacktriangleright$ {\bf (N-I) $ A_2= - A_1 \neq - G/X$} \\

The second and third conditions can be solved for $A_4$ and $A_5$ if $A_1 \neq G/X$:
\bea
\label{A4}
A_4&=& \frac{1}{8(G - A_1 X)^2}\left[ 4 G \left(3 (A_1 - 2 G_X)^2 - 2 A_3 G\right) -A_3 X^2 (16 A_1 G_X + A_3 G)  \right. \nb \\
&& \left. \qquad\qquad\qquad + 4 X \left(3 A_1 A_3 G + 16 A_1^2 G_X - 16 A_1 G_X^2 - 4 A_1^3 + 2 A_3 G G_X\right) \right] \,,  \\
\label{A5}
A_5 &=& 
\frac{1}{8 (G - A_1 X)^2}(2 A_1 - A_3 X - 4 G_X) \left[A_1 (2 A_1 + 3 A_3 X - 4 G_X) - 4 A_3 G\right] \,.
\eea
$A_1$  and $A_3$ are free functions thus there are three free functions 
($G, A_1$ and $A_3$). The solution for $a$ and $b$ is given by 
\be
b = \frac{2 G A_3 - A_1 (2 A_1 + A_3 X - 4 G_X)}{(G - A_1 X) (2 A_1 - A_3 X - 4 G_X)} \, a. 
\ee

The combination of Horndeski and beyond Horndeski theories is included in this class of models with 
\begin{equation}
A_2= - A_1 = - 2 G_X + X F \,, \qquad A_3 = - A_4 = - 2F \,, \qquad  A_5=0 \,. 
\label{HBH}
\end{equation}
However we can still add a free function $A_3$ with $A_4$ and $A_5$ satisfying Eqs.~(\ref{A4}) and (\ref{A5}) without spoiling the existence of the primary constraint. Thus this class of theories provide an extension of Horndeski plus beyond Horndeski with one more free function. We can take $G(X) \to 0$ limit smoothly and Eqs.~(\ref{A4}) and (\ref{A5}) become 
\be
A_4= - \frac{2 A_1}{X}\,, \qquad 
A_5=  \frac{(2 A_1 - A_3 X) (2 A_1 + 3 A_3 X)}{8 A_1 X^2} \,.
\label{NIG0}
\ee
The solution for $A_5$ agrees with (\ref{MIa5}) with $A_2=-A_1$, thus the limit $G(X) \to 0$ gives the subclass of (M-I) with $A_2=-A_1$. 

Notice that for the following choice of parameters
\begin{equation}
A_3 = \frac{2 (A_1 - 2 G_X)}{X}\,, 
\end{equation}
we have $a=0$ and $A_4= -A_3\,,  A_5=0$. If also $A_1= 2 G_X$, this is nothing but Horndeski and $a$ and $b$ are undetermined. In this case $V_*$ does not appear in the Lagrangian thus the primary constraint is simply given by $\pi_{*} \approx 0$. 

\hspace{1cm}\\
$\blacktriangleright$ {\bf (N-II) $A_2= - A_1 = - G/X$} \\

In this case the two conditions reduce to 
\begin{equation}
A_3 =  \frac{2\left(G - 2 X G_{X}\right)}{X^2} \,,
\end{equation}
and there are no constraints on $A_4$ and $A_5$. Thus there are three free functions ($G, A_4$ and $A_5$). The solution for $a$ is given by $a=0$.
 
\hspace{1cm}\\
$\blacktriangleright$ {\bf (N-III) $A_1 + A_2 \neq 0$} \\

We can solve the first condition for $A_4$
\begin{equation}
A_4 = \frac{2 G}{X^2} + \frac{8 G_X^2}{G} - \frac{2 (A_1 + 2 G_X)}{X} \,,
\label{a4}
\end{equation}
and then solve $A_5$ using Eq.~(\ref{third}). Substituting $A_4$ and $A_5$ into Eq.~(\ref{second}), we obtain
\begin{equation}
(G - A_1 X) \left[4 G^2 - 4 (A_1 + 3 A_2) G_X X^2 + 
G X (2 A_1 + 8 A_2 - 4 G_X + A_3 X)\right]=0 \,.
\end{equation}
Hence there are two branches of solutions. 

\begin{enumerate}[(i)]
\item $A_1 \neq G/X$

In this case we can solve $A_3$ in terms of $A_1$ and $A_2$:
\bea
A_3 &=& \frac{4 G_X (A_1 + 3 A_2)}{G} - \frac{2 (A_1 + 4 A_2 - 2 G_X)}{X} - \frac{4 G}{X^2} \,, \\
A_5 &=& \frac{2}{G^2 X^3} \left[4 G^3 + G^2 X (3 A_1 + 8 A_2 - 12 G_X) \right. \nb \\
&& \left. \qquad\quad + 8 G\, G_X X^2 (G_X - A_1 - 3 A_2)+6 G_X^2 X^3 (A_1 + 3 A_2)\right] \,.
\eea
Thus there are three free functions in this case ($G, A_1$ and $A_2$).
The solution for $a$ and $b$ is given by 
\be
b = - \frac{2\left(G - X G_X \right)}{X\left( G - 2 X G_X\right)} \, a
\ee

\item $A_1=G/X$

Here we find 
\bea
A_4 &=&\frac{8 G_X^2}{G} - \frac{4 G_X}{X} \,, \\
A_5 &=& \frac{1}{4 G X^3 (G + A_2 X)} \left[ G A_3^2 X^4 - 4 G^3 - 8 G^2 X (A_2 - 2 G_X)
 \right. \nb \\
&& \left. - 4 G X^2 \left(4 G_X \left(G_X - 2 A_2\right) + A_3 G\right) + 8 G_X X^3 (A_3 G - 4 A_2 G_X) \right] \,,
\eea
and there are three free functions ($G,\, A_2$ and $A_3$). The solution for $a$ and $b$ is given by 
\be
b = \frac{4 X (G + A_2 X) (X G_X - G) + A_*^2 \left[2 G^2 - 8 A_2 G_X X^2 + G X (4 A_2 + A_3 X - 4 G_X) \right]}{2 X \left(X + A_*^2 \right) (G + A_2 X ) (G - 2 G_X X)} \, a
\ee 

\end{enumerate}
	
\section{Conformal and disformal transformation}

\no After classifying the conditions for obtaining a primary constraint within EST theories, we now investigate  whether these set-ups
can be obtained from known Lagrangians through conformal and disformal transformations. First
we identify the class of theories minimally coupled with gravity (i.e. $G=0$)  that can be obtained from beyond Horndeski (\ref{BH}) by a conformal transformation.
Then,   we  study the class of theories that can be obtained from Horndeski theory (\ref{H}) by a conformal and disformal transformation together. 

\subsection{Conformal transformation on Beyond Horndeski} 

\no It was shown that under the generalised disformal transformation 
\begin{equation}
\bar{g}_{\mu \nu} = g_{\mu \nu} + \Gamma(X) \phi_{\mu} \phi_{\nu}. 
\label{disconf}
\end{equation}
beyond Horndeski theory is transformed to itself:
\begin{equation}
\bar{L}_{BH}[\bar{F}] = L_{BH}[F], 
\end{equation}
where $F= \bar{F}/(1+ X \Gamma)^{5/2}$. On the other hand, under the generalised conformal transformation
\begin{equation}
\bar{g}_{\mu \nu} = \Omega(X) g_{\mu \nu},
\end{equation}
it transforms as 
\begin{equation}
\bar{L}_{BH}[\bar{F}] = L_{BH}[F] + L_3[A_3] + L_5[A_5], 
\end{equation}
where 
\begin{equation}
F= \frac{\bar{F}}{\Omega}\,, \qquad A_3= \frac{4\,\bar{F}\, X \, \Omega_X}{\Omega^2}\,,
\qquad A_5 = \frac{2\,\bar{F}\, \Omega_X \left( 3\,X\, \Omega_X - 2\,\Omega \right)}{\Omega^3}\,.
\end{equation}
In terms of $A_i$, this gives 
\bea
&&A_2= -A_1 =X F\,, \qquad  A_3= - 2 F+ \frac{4\,F\, X \, \Omega_X}{\Omega}\,, \nb \\
&& A_4=  2 F\,, \qquad  
A_5 = \frac{2\,F\, \Omega_X \left( 3\,X\, \Omega_X - 2\,\Omega \right)}{\Omega^2}\,. 
\label{BHconf}
\eea
These satisfy $A_4= - 2 A_1/X$ and (\ref{MIa5}). Thus this theory is included in the case (M-I). This is indeed the limit of $G(X) \to 0$ in (N-I) given by Eq.~(\ref{NIG0}).

\subsection{Conformal and disformal transformation on Horndeski}

\no The conformal and disformal transformation 
\begin{equation}
\bar{g}_{\mu \nu} = \Omega(X) g_{\mu \nu} + \Gamma(X) \phi_{\mu} \phi_{\nu}
\end{equation}
transforms the Horndenski action as 
\begin{equation}
\bar{L}_H[\bar{G}] = L_H[G] + L_{BH}[F] + L_3[A_3] +  L_4[A_4] + L_5[A_5]
\label{confdis}
\end{equation}
where
\bea
G &=&   \bar{G} \sqrt{\Om\left(\Om + X \Ga\right)} \,, \\[1ex]
F &=& \frac{\bar{G} \left[X \Omega \Gamma_X + (2 \Omega + X \Gamma ) \Omega_X \right]}{X \sqrt{\Omega (\Omega + X \Gamma )}} - \frac{2 \bar{G}_{\bar{X}} \sqrt{\Omega } \left(\Omega_X + X \Gamma_X \right)}{(\Omega + X \Gamma )^{3/2}} \,,\\[1ex]
A_3 &=&  \frac{4 \sqrt{\Omega }\, \Omega_X \left[\bar{G} (\Omega + X \Gamma ) - 2 X \bar{G}_{\bar{X}} \right]}{X (\Omega + X \Gamma )^{3/2}} \,, \\[1ex]
A_4 &=& \frac{2 \bar{G} \Omega_X \left[X (3 \Omega + X \Gamma  ) \Omega_X - 2 \Omega \left(\Omega - X^2  \Gamma_X \right)\right]}{X \Omega^{3/2} \sqrt{\Omega + X \Gamma }} - \frac{8 \bar{G}_{\bar{X}} \Omega \,\Omega_X \left[X \left(\Omega_X + X \Gamma_X \right) - \Omega \right]}{[\Omega (\Omega + X \Gamma )]^{3/2}} \,, \\[1ex]
A_5 &=& - \frac{2 \bar{G} \Omega_X \left(2 \Omega \Gamma_X + \Gamma \Omega_X \right)}{\Omega^{3/2} \sqrt{\Omega + X \Gamma }} - \frac{4 \bar{G}_{\bar{X}} \Omega \, \Omega_X \left(\Omega_X - 2 X \Gamma_X \right)}{[\Omega (\Omega + X \Gamma )]^{3/2}} \,.
\eea
We can check that this theory satisfies the conditions $A_2= - A_1$ and Eqs.~(\ref{A4}) and (\ref{A5}). Thus this theory is included in case (N-I). Theories in case (N-I) have three free functions. On the other hand, the action (\ref{confdis}) contains $\bar{G}$, $\Omega$ and $\Gamma$. Thus there is the same number of free functions. Indeed we can relate $\bar{G}$, $\Omega_X$ and $\Gamma_X$ to $G$, $F$ and $A_3$ as 
\bea
\bar{G} &=& \frac{G}{\sqrt{\Omega(\Omega + X \Gamma)}} \,,\\[1ex]
\Omega_X &=& \frac{A_3 X \Omega}{4 \left(X^2 F + G - 2 X G_X\right)} \,, \\[1ex]
\Gamma_X &=& \frac{\Omega \left[ 2 G (2 F - A_3) + X \left(F X - 2 G_X \right)(4 F - A_3)\right]}{4 \left[G + X \left(F X - 2 G_X\right) \right]^2} \,.
\eea
Thus the theories in case (N-I) can be mapped to Horndeski if the transformation (\ref{confdis}) is invertible. Note that the transformation is not always invertible. In fact, in Ref.~\cite{Crisostomi:2016tcp}, we showed that beyond Horndeski theory cannot be mapped to Horndeski as $\Omega=1$ and $\Gamma=- 1/X$ and the transformation is indeed not invertible.

\section{Phenomenological consequences of EST theories}

\no In this section we discuss possible phenomenological applications of the EST theories introduced in \cite{Langlois:2015cwa}. We first study whether these theories admit a healthy Minkowski limit; then we discuss their implications for the effective field theory of dark energy around the Friedmann-Robertson-Walker (FRW) background. 

\subsection{Minkowski limit}

\no We consider the Minkowski limit by taking $g_{\mu \nu} = \eta_{\mu \nu}$. Notice that we do not consider any fluctuations of metric, thus by definition the extrinsic curvature vanishes. In order to have a primary constraint we need therefore to impose the condition $\pi_* \approx 0$. For general combinations of ${\cal L}_{i}[A_i]$, $\pi_*$ is given by 
\begin{equation}
\pi_* = 2\left[ A_1 + A_2  - A_*^2 (A_3 + A_4) + A_*^4 A_5 \right] V_* \,.
\end{equation}
In order to have Lorentz invariant solutions for $A_i(X)$, we need to impose the following conditions \cite{Langlois:2015cwa}: 
\begin{equation}
A_2= -A_1\,, \qquad A_3= -A_4\,, \qquad A_5=0 \,. 
\label{flatcon}
\end{equation}
This is   a generalised galileon described by the following actions:
\begin{equation}
S_{\text{gal}} = \int d^4 x \, M(X) \left[  \phi_{\mu \nu} \phi^{\mu \nu}
- (\Box \phi)^2 \right] 
 =  \int d^4 x \, 2 M_X(X) \left[ (\Box \phi) \phi^{\mu} \phi_{\mu \nu} \phi^{\nu} - \phi^{\mu} \phi_{\mu \rho} \phi^{\rho \nu} \phi_{\nu} \right],
\end{equation}
where we used the fact that $\phi_{\mu \nu} \phi^{\mu \nu}
- (\Box \phi)^2$ is a total derivative in the Minkowski spacetime. 

We summarise below whether each class of theories identified in previous sections can satisfy the condition (\ref{flatcon}) or not. 

\begin{itemize}
	\item (M-I):  Beyond Horndeski satisfies (\ref{flatcon})
	\item  (M-II): No theory can satisfy (\ref{flatcon}). 
	\item  (M-III): A theory with $A_1=A_2=0$, $A_3= - A_4$, $A_5=0$ satisfies (\ref{flatcon}).
	\item (N-I): Beyond Horndeski and Horndeski satisfy (\ref{flatcon}). 
	\item (N-II): By chosing $A_3= - A_4$, $A_5=0$, it is possible to satisfy  (\ref{flatcon}).  
	\item{(N-III)}: No theory can satisfy (\ref{flatcon}).
\end{itemize}

This shows that beyond Horndeski and Horndeski are not the only theories that propagate three degrees of freedom on curved spacetime in vacuum and have a healthy Minkowski space limit. On the other hand, theories that do not satisfy (\ref{flatcon}) could have a healthy decoupling limit around a non-trivial background such as FRW.

\subsection{Unitary gauge and effective field theory of dark energy}

\no Around the FRW background, for linear perturbations, we can choose a gauge where the scalar field only depends on time, $\phi= \phi(t)$. Using the time reparametrisation, we can further impose the condition $\phi =t$. This is known as unitary gauge. In the literature, the unitary gauge is frequently used to study extended theories of Horndeski \cite{Gleyzes:2014dya, Gleyzes:2014qga, Gao:2014fra}. Although care must be taken to draw conclusions on the number of degrees of freedom using this gauge \cite{Deffayet:2015qwa}, once this issue has been clarified in a gauge invariant way, the unitary gauge is particularly useful to study linear cosmological perturbations and observational consequences of the models. 

In the unitary gauge $\phi=t$, $\hat{A}=0$ and the Lagrangian densities reduce to 
\bea
{\cal L}_1 &= & A_1 \left(A_*^2 K_{\mu \nu} K^{\mu \nu} + V_*^2 \right)\,,\\
{\cal L}_2 &= & A_2 \left( A_* K + V_* \right)^2 \,, \\
{\cal L}_3 &= & - A_3  A_*^2 V_* \left(A_* K +  V_*\right)\,,  \\
{\cal L}_4 &= & -A_4 A_*^2 V_*^2\,, \\
{\cal L}_5 &= & A_5 A_*^4 V_*^2\,,  \\
{\cal L}_R &= & G \left( K_{\mu \nu} K^{\mu \nu} - K^2 \right) + 4 G_X A_* V_* K \,,
\eea
and $X=-A_*^2$. New contributions in the extended theories compared with Horndeski and beyond Horndeski theories come from $V_*$:
\begin{equation}
{\cal L} =  I(A_*) V_*^2  + J(A_*) K V_*,
\end{equation}
where 
\begin{equation}
I(A_*) = A_1 + A_2 - A_*^2 \left(A_3 + A_4\right) + A_*^4 A_5\,, \qquad 
J(A_*) = A_* \left(2 A_2 - A_*^2 A_3 + 4 G_X\right)\,.
\end{equation}
Horndeski and beyond Horndeski gives $I(A_*)=0$ and $J(A_*)$=0 but the EST theories in general give non-zero $I(A_*)$ and $J(A_*)$. In the unitary gauge $A_* = N$ therefore, using (\ref{Vstar}), $V_*= (\dot{N} -  N^{i} \partial_i N)/N$. These operators have not been considered so far in the so-called effective field theory of dark energy \cite{Gubitosi:2012hu,Bloomfield:2012ff}, which starts from a general Lagrangian in the unitary gauge \cite{Gleyzes:2013ooa}:
\begin{equation}
{\cal L} = L(N, K, K_{\mu \nu} K^{\mu \nu}, {}^{(3)} R, {}^{(3)} R_{\mu \nu} K^{\mu \nu}, {}^{(3)} R_{\mu \nu} {}^{(3)} R^{\mu \nu} ) \label{EFTL}
\end{equation}
The dependence on $(\dot{N} -  N^{i} \partial_i N)$ is normally ignored a priori as the lapse could become dynamical and lead to a dangerous additional mode\footnote{Ref.~\cite{Gleyzes:2014rba} pointed out that the appearance of $\dot{N}$ does not necessarily lead to an additional propagating degree of freedom since the general conformal and disformal transformation (\ref{disconf}) can remove the $\dot{N}$ dependence.}. However, as we showed, there are healthy theories that contain this operator in the unitary gauge. Thus the general Lagrangian density (\ref{EFTL}) can include the following new operators 
\begin{equation}
(\dot{N} -  N^{i} \partial_i N)^2\,, \qquad (\dot{N} -  N^{i} \partial_i N) K \,. 
\end{equation}
These operators break the symmetry $t \to \tilde{t}(t)$ and we are left only with the invariance under time translations $\tilde{t} = t\, +$ const. \cite{Blas:2010hb}. It would be interesting to study phenomenological consequences of these operators. 

\section{Summary and Discussion}

\no We studied a large class of Extended Scalar-Tensor (EST) theories of gravity recently introduced in \cite{Langlois:2015cwa}, which are at most quadratic in second derivatives of the scalar field, and propagate at most three degrees of freedom in the vacuum. Despite the presence of higher derivatives, these theories are characterised by a primary constraint which is a necessary condition to forbid the propagation of an additional dangerous dof. We derived the conditions for the existence of a primary constraint, which are equivalent to the degeneracy conditions for the kinetic matrix obtained in \cite{Langlois:2015cwa}, and classified in full generality solutions for these conditions. In addition, we identified EST theories that could be mapped from Horndeski or beyond Horndeski theories by generalised conformal and disformal transformations. Finally, we explored interesting consequences of EST theories for gravity and cosmology. We investigated which ones among  the new theories admit a healthy Minkowski limit, a necessary condition to apply EST theories to weakly gravitating systems. Then, we  examined the possible relevance of EST theories to dark energy, using the language  of the Effective Field Theory (EFT).  We showed  that  these theories can be associated with novel EFT operators of dark energy, which are absent in Horndeski and beyond Horndeski theories. Hence, these theories can have potentially distinctive observational consequences for dark energy. Table \ref{tableI} summarises the properties of the EST theories of gravity we determined. 

\begin{table}[h]
\begin{tabular}{ |p{3cm}||p{3cm}|p{3cm}|p{5cm}|  }
 		\hline
 		\multicolumn{4}{|c|}{Minimally coupled theories} \\
 		\hline
 		Classification & Free functions & Minkowski limit &Examples\\
 		\hline
 		M-I   &  3    & \checkmark\, (BH)  & BH, EBH${}^{(1)}$, BH+$\Omega$${}^{(2)}$  \\
 		M-II  &  3  & X & \\
 		M-III & 4  &  \checkmark  &  ${\cal L}_i (i=2,3,4,5)$ \\ 
 		\hline
 	\end{tabular}	
 	\begin{tabular}{ |p{3cm}||p{3cm}|p{3cm}|p{5cm}|  }
 		\hline
 		\multicolumn{4}{|c|}{Non-minimally coupled theories} \\
 		\hline
 		Classification & Free functions & Minkowski limit &Examples\\
 		\hline
 		N-I   &  3    & \checkmark\, (H, BH) & H, H+$\Gamma$ (H+BH)${}^{(3)}$, H+$\Omega$+$\Gamma$${}^{(4)}$ \\
 		N-II  & 3  &  \checkmark  &  \\ 
 		N-III (i) & 3 & X & \\
 		N-III  (ii)  & 3 & X &  \\
 		\hline
 	\end{tabular}
 \caption{Summary of Extended Scalar-Tensor (EST) theories. $(1)$: Extension of Beyond Horndeski theory given by Eq.~(\ref{bbh-1}). $(2)$: Theories obtained by the conformal transformation from Beyond Horndeski. $(3)$: Theories obtained by the disformal transformation from Horndeski. This is equivalent to a combination of Horndeski and beyond Horndeski. $(4)$: Theories obtained by the conformal and disformal transformation from Horndeski. }\label{tableI}
 \end{table}
   
Our analysis opens up new perspectives for building scalar-tensor theories of gravity, with potentially interesting applications for inflation and dark energy. Our method to find conditions for the existence of primary constraints can be straightforwardly applied to study theories that are higher than quadratic in second derivatives of the scalar fields, which nevertheless have
primary constraints that prevent the propagation of additional dangerous modes. While  theories that are cubic in second derivatives extend quintic beyond Horndeski, systems that are higher than cubic lead to completely new theories, which deserve to be explored. 
   
A particular simple example of such theories can be found extending the Lagrangian we presented in Section \ref{sub-mct} (see the discussion in Eq~\eqref{bbh-1} and below) in terms of projection tensors:
  \be
  {\cal L}_n\,=\,{\cal N}_{\mu_1\,\dots\,\mu_n}^{\alpha_1\dots \alpha_n}\,\phi_{\alpha_1}^{\,\mu_1}\,\dots\,\phi_{\alpha_n}^{\,\mu_n}
   \ee
with the tensor ${\cal N}_{\mu_1\,\dots\,\mu_n}^{\alpha_1\dots \alpha_n}$ being an arbitrary combination of
$P_{\mu}^{\alpha}$ as
\be
{\cal N}_{\mu_1\,\dots\,\mu_n}^{\alpha_1\dots \alpha_n}\,=\,Q_1(\phi, {X})\,P_{\mu_1}^{\alpha_1}\,P_{\mu_2}^{\alpha_2}\,\dots \,P_{\mu_n}^{\alpha_n}+
Q_2(\phi, {X})\,P_{\mu_2}^{\alpha_1}\,P_{\mu_1}^{\alpha_2}\,\dots \,P_{\mu_n}^{\alpha_n}+\dots
\ee
where we can include all the permutations in the lower indexes of the projectors. For the very same reasons explained in Section \ref{sub-mct}, such Lagrangians enjoy an extra primary constraint. 

Given these considerations, a natural question is whether or not  there exists a {closed form} for the {\it most general } scalar-tensor Lagrangian propagating three degrees of freedom, containing arbitrary high derivatives. Given its relevance for cosmological model building, we aim to answer this question in a forthcoming paper. 

\section*{Acknowledgments}

\no We thank Matthew Hull for discussions. We also thank David Langlois and Karim Noui for useful correspondence. KK is supported by the UK Science and Technology Facilities Council grants ST/K00090X/1 and the European Research Council grant through 646702 (CosTesGrav).
GT is partially supported by the STFC grant ST/N001435/1.

\end{document}